\documentstyle[aps,epsf,floats]{revtex}

\epsfverbosetrue    
\def\postscript#1{\begin{center}\leavevmode \hbox{\epsfxsize=\columnwidth\epsfbox{#1}}\end{center}}

\begin{document}

\twocolumn[\hsize\textwidth\columnwidth\hsize\csname@twocolumnfalse\endcsname

\title{Effects of Interlayer Interaction on the Superconducting State in YBCO}
\author{C. O'Donovan\cite{me} and J. P. Carbotte}
\address{Department of Physics \& Astronomy, McMaster University, 
Hamilton, Ontario, Canada L8S 4M1}
\date{October 16, 1996}
\maketitle

\begin{abstract}
For a two layer system in a weak coupling {\sc bcs} formalism any
interlayer interaction, regardless of its sign, enhances the critical
temperature. The sign has an effect upon the relative phase of
the order parameter in each of the two planes but not upon its
magnitude. When one of the planes has a dispersion consistent with CuO
chains and no intrinsic pairing interaction there is both an
enhancement of the critical temperature and an $s+d$ mixing in both
layers as the interlayer interaction is increased.  The magnetic
penetration depth, $c$-axis Josephson tunneling, density of states
and Knight shift are calculated for several sets of model parameters.
\end{abstract}


]

\narrowtext

\section{Introduction}

The search for the mechanism which causes superconductivity in the
copper oxide materials is an ongoing effort which has yet to reach a
consensus. One factor which any model should account for is that the
critical temperature tends to be higher in systems with multiple
adjacent CuO$_2$ layers; and even in systems, such as YBCO, in which a
CuO layer is adjacent to the CuO$_2$ layers $T_{\rm c}$ seems to be
enhanced.

It is generally believed that the superconducting condensate resides
in the CuO$_2$ planes, although one interpretation of the observed
large $x$-$y$ anisotropy of the zero temperature magnetic penetration
depth (a factor of $\sim1.6$) in YBCO indicates that there is as much
condensate in the CuO chains as in the CuO$_2$ planes (ie, the
condensate in the chains only contributes to the penetration depth for
current in the direction parallel to the chains, ie the
$b$-direction).\cite{basov} Since it is believed that whatever
mechanism is responsible for superconductivity in the copper oxides is
intrinsic to the CuO$_2$ planes some other mechanism for creating
superconducting condensate on the CuO chains is required.

In this paper we derive a Hamiltonian for a layered system and, making
a simplifying assumption that there is no pairing between electrons which
reside in different layers, derive a pair of coupled {\sc
bcs} equations for a system of two layers, each with possibly
different dispersion and pairing interactions.  In this model Cooper
pairs can scatter between the layers so that, as in a two band
model,\cite{mmp} even if there is no pairing interaction in one of the
layers there will still be a condensate in that layer due to the
interlayer interaction.

Although we make a particular choice for the pairing interaction
(which is motivated by the nearly antiferromagnetic Fermi liquid model
which leads naturally to a $d$-wave gap for single CuO$_2$
planes.\cite{chi}) a $d$-wave solution is also found for other types
of non-isotropic pairing interactions.\cite{odonovan1} One result of
having gap nodes cross the Fermi surface is a low temperature linear
behaviour of the magnetic penetration depth, $\lambda_{ii}^{-2}$, as
is observed in YBCO.\cite{basov,bonn,hardy2,bonn2} Here we only try to
model the low temperature behaviour and relative magnitude\cite{basov}
in the $x$ and $y$ directions of the magnetic penetration depth. The
$T\sim T_{\rm c}$ behaviour is only reproduced for values of
$2\mit\Delta_{\rm max}/T_{\rm c}$ about $1.5$ times
higher\cite{scalapino} than the value of 4.4 found in the {\sc bcs}
weak coupling approximation. In the two layer model that we study here
a higher value of $2\mit\Delta_{\rm max}/T_{\rm c}$ is obtained which
is closer to that observed in experiments such as {\sc
arpes}\cite{ding} or current-imaging tunneling spectroscopy ({\sc
cits})\cite{edwards} that measure the absolute magnitude of the order
parameter.

We find that the presence of the chains destroys the tetragonal
symmetry of the CuO$_2$ planes and shifts the $d$-wave gap nodes in
the CuO$_2$ plane off the diagonals in agreement with an earlier
model.\cite{odonovan3} In this case the gap contains an admixture of
$s$- and $d$-wave symmetry. Calculations of the $c$-axis Josephson
tunneling current show that the positive and negative parts of the
order parameter do not cancel, as for $d$-wave pairing in a tetragonal
system, and that the Josephson junction resistance-tunneling current
products, $RJ(T=0)$, are in the range of 0.1-3.0meV, in agreement with
the experiments of Sun {\it et al},\cite{sun}.  We also calculate both
the normal and superconducting density of states ({\sc dos}) for the
CuO$_2$ planes and CuO chains separately since some surface probes,
such as {\sc cits},\cite{edwards} can measure them separately. Finally
the Knight shift is calculated separately for both the planes and the
chains.

In section II we introduce our Hamiltonian and derive a set of coupled
{\sc bcs} equations for planes and chains and other necessary
formulas, particularly the expression for the magnetic penetration
depth in this model. In section III we present the the solutions of
these {\sc bcs} equations as well as the results of calculations of
the magnetic penetration depth, electronic density of states, $c$-axis
Josephson tunneling, and Knight shift. Section IV contains a short
discussion and conclusion.

\section{Formalism}

In this section we will present a Hamiltonian in which multiple layers
are coupled through the pairing interactions between adjacent layers.
We will then make the assumption that there is no interlayer pairs
(ie, that each Cooper pair resides in only one of the layers) and that
there is no single particle interlayer hopping. The Hamiltonian of
this special case is then diagonalized and coupled {\sc bcs} equations
derived. We then give expressions for the magnetic penetration depth
in this model, the Knight shift, the quasi-particle density of states
and the DC Josephson junction resistance-tunneling current product
for a $c$-axis tunnel junction.

The general Hamiltonian is:

\begin{eqnarray}
H =&&\sum_{{\bf k},\alpha\beta}{
\varepsilon_{{\bf k},\alpha\beta}
\left(a_{{\bf k}\uparrow,\alpha}^\dagger a_{{\bf k}\uparrow,\beta}
+a_{{\bf k}\downarrow,\beta}^\dagger a_{{\bf k}\downarrow,\alpha}\right)
} \nonumber \\ 
&-&\frac{1}{\Omega} \sum_{{\bf k},{\bf q},\alpha\beta\gamma\delta}{
V_{{\bf k},{\bf q},\alpha\beta\gamma\delta}a_{{\bf
 k}\uparrow,\alpha}^\dagger a_{-{\bf k}\downarrow,\beta}^\dagger a_{{\bf q}\uparrow,\gamma}
 a_{-{\bf q}\downarrow,\delta} },
\label{hamilton.eq}
\end{eqnarray}

\noindent where the greek indices enumerate the layers, the $a_{{\bf
k},\alpha}^\dagger$ ($a_{{\bf k},\alpha}$) create (destroy) electrons
of momentum ${\bf k}$ in layer $\alpha$ ({\bf k} is in units of
$a^{-1}$ where $a$ is the lattice parameter), $\varepsilon_{{\bf
k},\alpha\beta}$ is the electron dispersion, and $V_{{\bf k},{\bf
q},\alpha\beta\gamma\delta}$ is the pairing interaction.

Performing a mean field analysis, we get:

\begin{eqnarray*}
H =&&\sum_{{\bf k},\alpha\beta}{\varepsilon_{{\bf k},\alpha\beta}
\left(a_{{\bf k}\uparrow,\alpha}^\dagger a_{{\bf k}\uparrow,\beta}
+a_{{\bf k}\downarrow,\beta}^\dagger a_{{\bf k}\downarrow,\alpha}\right)
} \nonumber \\ 
 &-&\sum_{{\bf k},\alpha\beta}{\left(\mit\Delta_{{\bf k},\alpha\beta}a_{{\bf
 k}\uparrow,\alpha}^\dagger a_{-{\bf k}\downarrow,\beta}^\dagger+ {\rm H.c.}\right)}+C,
\end{eqnarray*}

\noindent where $C$ is a constant, H.c.\ indicates the Hermitian
conjugate, $\mit\Delta_{{\bf k},\alpha\beta}\equiv
\Omega^{-1}\sum_{{\bf q},\gamma\delta}{V_{{\bf k},{\bf
q},\alpha\beta\gamma\delta}\chi_{{\bf q},\gamma\delta}}$ are the order
parameters and $\chi_{{\bf q},\gamma\delta}\equiv
\left<a_{{\bf q}\uparrow,\gamma} a_{-{\bf q}\downarrow,\delta}\right>$
are the pair susceptibilities.

Writing this in Nambu spinor notation, we get:

\begin{eqnarray*}
H=\sum_{{\bf k},\alpha\beta}A_{{\bf k},\alpha}^\dagger \hat H_{{\bf
k},\alpha\beta}A_{{\bf k},\beta}
\end{eqnarray*}

\noindent where $A_{{\bf k},\alpha}^\dagger\equiv
\left[\begin{array}{cc}
a_{{\bf k}\uparrow,\alpha}^\dagger & a_{-{\bf k}\downarrow,\alpha}
\end{array}\right]$
and:

\begin{equation}
\hat H_{{\bf k},\alpha\beta}\equiv\left[ \begin{array}{cc} \varepsilon_{{\bf
k},\alpha\beta} & \mit\Delta_{{\bf k},\alpha\beta} \\
\mit\Delta_{{\bf k},\alpha\beta}^\dagger & -\varepsilon_{{\bf k},\alpha\beta}
\end{array}\right].
\label{hamiltonian2.eq}
\end{equation}

In this model the magnetic penetration depth is given by the
expression:\cite{atkinson}

\begin{eqnarray}
\label{pd1.eq}
\lambda_{ij}^{-2}&&=\frac{4\pi e^2}{\hbar^2c^2}\frac{1}{\Omega}\sum_{{\bf k},\alpha\beta}
\hat\gamma^{(i)}_{{\bf k},\alpha\beta}\hat\gamma^{(j)}_{{\bf k},\beta\alpha}
\left(\hat G_{{\bf k},\alpha\beta}\big|_{\mit\Delta=0}-\hat G_{{\bf k},\alpha\beta}\right)
\end{eqnarray}

\noindent where:

\begin{eqnarray}
\hat G_{{\bf k},\alpha\beta}\equiv&&\frac{\partial f(E_{{\bf k},\alpha})}{\partial E_{{\bf k},\alpha}}\delta_{\alpha\beta}
\nonumber\\
&+&\frac{f(E_{{\bf k},\alpha})-f(E_{\bf k,{\beta}})}{E_{{\bf k},\alpha}-E_{\bf k,{\beta}}}(1-\delta_{\alpha\beta}),\nonumber\\
\hat\gamma^{(i)}_{{\bf k},\alpha\beta}\equiv&&\sum_{\gamma\delta}\hat U_{{\bf k},\alpha\gamma}^\dagger\frac{\partial \varepsilon_{{\bf k},\gamma\delta}}
{\partial k_i}\hat U_{{\bf k},\delta\beta},\nonumber
\end{eqnarray}

\noindent $\pm E_{{\bf k},\alpha}$ are the eigenvalues of
Eq.~(\ref{hamiltonian2.eq}), $f(x)$ is the Fermi function,
$\delta_{\alpha\beta}$ is a Kronecker delta and $\hat U_{{\bf
k},\alpha\beta}$ is the unitary matrix which diagonalizes
Eq.~(\ref{hamiltonian2.eq}). The quantities $e$, $\hbar$, and $c$ are
the electron charge, Planck's constant and the speed of light
respectively.

We now make the following simplification: $V_{{\bf k},{\bf
q},\alpha\gamma} \equiv V_{{\bf k},{\bf q},\alpha\alpha\gamma\gamma} =
V_{{\bf k},{\bf
q},\alpha\beta\gamma\delta}\delta_{\alpha\beta}\delta_{\gamma\delta}$
and $\varepsilon_{{\bf k},\alpha}\equiv\varepsilon_{{\bf
k},\alpha\alpha}=\varepsilon_{{\bf
k},\alpha\beta}\delta_{\alpha\beta}$. This means that there is only
intralayer pairing and no interlayer pairing (ie, $\mit\Delta_{{\bf
k},\alpha}\equiv\mit\Delta_{{\bf k},\alpha\beta}\delta_{\alpha\beta}$
and $\chi_{{\bf q},\gamma}\equiv\chi_{{\bf
q},\gamma\delta}\delta_{\alpha\beta}$ are both diagonal in the greek
indices) and there is no single particle interlayer hopping. This
Hamiltonian has the same form as that for a two band model studied by
one of us \cite{chi} in an earlier publication and is similar to that
studied by others.\cite{klemm,klemm2,combescot} Interlayer
pairing\cite{combescot,kettemann,wheatley} has also been studied.

The Hamiltonian has eigenvalues given by $E_{{\bf k},\alpha}=\pm\sqrt{
\varepsilon_{{\bf k},\alpha}^2+\mit\Delta_{{\bf k},\alpha}^2}$ and is
diagonalized by the unitary matrix:

\begin{equation}
\label{unitary.eq}
\hat U_{{\bf k},\alpha}=\left[ \begin{array}{cc} u_{{\bf
k},\alpha} & v_{{\bf k},\alpha} \\
-v_{{\bf k},\alpha} & u_{{\bf k},\alpha}
\end{array}\right],
\end{equation}

\noindent where:

\begin{eqnarray}
u_{{\bf k},\alpha}&\equiv&\sqrt{\frac{1}{2}
\left(1+\frac{\varepsilon_{{\bf k},\alpha}}{E_{{\bf k},\alpha}}\right)} \nonumber\\
v_{{\bf k},\alpha}&\equiv&\sqrt{\frac{1}{2}
\left(1-\frac{\varepsilon_{{\bf k},\alpha}}{E_{{\bf k},\alpha}}\right)}\nonumber
\end{eqnarray}

\noindent are the usual {\sc bcs} coherence factors. Using this
unitary transformation (\ref{unitary.eq}) we can evaluate the pair
susceptibilities to get:

\begin{eqnarray}
\label{sus.eq}
\chi_{{\bf q},\alpha}&\equiv&
\left<a_{{\bf q}\uparrow,\alpha} a_{-{\bf q}\downarrow,\alpha}\right> \nonumber \\
&=&\frac{\mit\Delta_{{\bf q},\alpha}}{2E_{{\bf q},\alpha}}
\tanh \left( \frac{E_{{\bf q},\alpha}}{2 k_{\rm B}T} \right),
\end{eqnarray}

\noindent 
where $T$ is the temperature and $k_{\rm B}$ is Boltzmann's
constant. Note that if we had included the interlayer pairing from
Eq.~\ref{hamilton.eq} we would have susceptibilities of the form
$\left<a_{{\bf q}\uparrow,\alpha} a_{-{\bf q}\downarrow,\beta}\right>$
with $\alpha\not =\beta$ and both the eigenvalues and the unitary
matrix (\ref{unitary.eq}) would be much more complicated.

For a bilayer system (ie, $\alpha=1,2$) the {\sc bcs} equations are:

\begin{eqnarray}
\label{bcs.eq}
\mit\Delta_{{\bf k},1}&=& \frac{1}{\Omega}\sum_{{\bf q}}{\left(
V_{{\bf k},{\bf q},11}\chi_{{\bf q},1}
+V_{{\bf k},{\bf q},12}\chi_{{\bf q},2}
\right)} \nonumber \\
\mit\Delta_{{\bf k},2}&=& \frac{1}{\Omega}\sum_{{\bf q}}{\left(
V_{{\bf k},{\bf q},12}\chi_{{\bf q},1}
+ V_{{\bf k},{\bf q},22}\chi_{{\bf q},2}
\right)},
\end{eqnarray}

\noindent where we have taken $V_{{\bf k},{\bf q},12}=V_{{\bf k},{\bf
q},21}$, although in general only $V_{{\bf k},{\bf q},12}=V_{{\bf k},{\bf
q},21}^\dagger$ is required.

Noting that $\chi_{{\bf q},2}$ changes sign (see Eq.~\ref{sus.eq}) with
$\mit\Delta_{{\bf k},2}$ we see that this set of equations
(\ref{bcs.eq}) is unchanged by the substitution $\{\mit\Delta_{{\bf
k},2},V_{{\bf k},{\bf q},12}\}\rightarrow
\{-\mit\Delta_{{\bf k},2},-V_{{\bf k},{\bf q},12}\}$ which means that the
overall sign of $V_{{\bf k},{\bf q},12}$ only affects the relative
sign of the order parameters in the two layers and not their
magnitudes. This is interesting because it means that the effect on
$T_{\rm c}$ of having an interlayer interaction is independent of
whether this interaction is attractive or repulsive, although some
calculated properties (eg, $c$-axis Josephson tunneling current) still
depend upon the relative sign of the interlayer interaction. {\em It
is important to emphasize that any interlayer interaction, either
attractive or repulsive, tends to enhance $T_{\rm c}$ and that this is
consistent with the observation that $T_{\rm c}$ is higher in
materials with multiple adjacent CuO layers.}  This well known result
can be easily shown by examining the coupled {\sc bcs} equations
(\ref{bcs.eq}) near $T\sim T_c$. In this limit we can write:
\begin{eqnarray*}
{\mit\Delta}_{{\bf k},\alpha} &=& {\mit\Delta}_\alpha \eta_{\bf k}\\
V_{{\bf k},{\bf q},\alpha\beta} &=& V_{\alpha\beta} \eta_{\bf k}\eta_{\bf q}
\end{eqnarray*}
where $\mit\Delta_\alpha$ and $V_{\alpha\beta}$ are numbers and
$\eta_{\bf k}$ is a normalized function which could be taken to be
$d$-wave and corresponds to the highest $T_c$. The coupled {\sc bcs}
equations (6) can then be written as:
\begin{mathletters}
\begin{eqnarray}
{\mit\Delta}_1 &=& 
{\mit\Delta}_1 V_{11}\frac{1}{\Omega}\sum_{{\bf q}}{
\frac{\left(\eta_{\bf q}\right)^2}{\varepsilon_{{\bf q},1}}
\tanh \left( \frac{\varepsilon_{{\bf q},1}}{2 k_{\rm B}T_c} \right)}\nonumber\\
&&+{\mit\Delta}_2 V_{12}\frac{1}{\Omega}\sum_{{\bf q}}{
\frac{\left(\eta_{\bf q}\right)^2}{\varepsilon_{{\bf q},2}}
\tanh \left( \frac{\varepsilon_{{\bf q},2}}{2 k_{\rm B}T_c} \right)}\\
{\mit\Delta}_2 &=& 
{\mit\Delta}_1 V_{12}\frac{1}{\Omega}\sum_{{\bf q}}{
\frac{\left(\eta_{\bf q}\right)^2}{\varepsilon_{{\bf q},1}}
\tanh \left( \frac{\varepsilon_{{\bf q},1}}{2 k_{\rm B}T_c} \right)}\nonumber\\
&&+{\mit\Delta}_2 V_{22}\frac{1}{\Omega}\sum_{{\bf q}}{
\frac{\left(\eta_{\bf q}\right)^2}{\varepsilon_{{\bf q},2}}
\tanh \left( \frac{\varepsilon_{{\bf q},2}}{2 k_{\rm B}T_c} \right)}.
\end{eqnarray}
\end{mathletters}

We now assume that $\mit\Delta_1$ is the dominant superconducting
channel when $V_{12}=V_{21}=0$ and obtain assuming an infinite band
with cutoff $\omega_C$:
\begin{mathletters}
\begin{eqnarray}
\mit\Delta_1 &=& \left(\mit\Delta_1\lambda_{11}+\mit\Delta_2\lambda_{12}\right)\ln\left(\frac{1.13\omega_C}{T_c}\right)
\label{silliness1}\\
\mit\Delta_2 &=& \left(\mit\Delta_1\lambda_{21}+\mit\Delta_2\lambda_{22}\right)\ln\left(\frac{1.13\omega_C}{T_c}\right),
\label{silliness2}
\end{eqnarray}
\end{mathletters}
where $\lambda_{ij}\equiv V_{ij}\times$ the density of electronic
states at the Fermi surface. Substitution of equation
(\ref{silliness2}) into (\ref{silliness1}) leads to a quadratic in
$\ln\left(1.13\omega_C/T_c\right)$ with solution:
\begin{equation}
T_c=1.13 e^{1/\tilde{\lambda}}
\end{equation}
with:
\begin{equation}
\tilde{\lambda}=\frac{1}{2}\left[\lambda_{11}+\lambda_{22}+\sqrt{(\lambda_{11}-\lambda_{22})^2+4\lambda_{12}\lambda_{21}}\right].
\end{equation}

This result is well known and is given by equation (6.3) on page 105
of Mechanisms of Conventional and High $T_c$
Superconductivity.\cite{kresin} It is also found as equation (40) of
H.\ Chi and Carbotte.\cite{chi} We note that $\lambda_{12}$, whatever
its sign, increases $\tilde{\lambda}$ and so increases $T_c$.

If we had taken $V_{{\bf k},{\bf q},12}$ as complex the symmetry would
be $\{\mit\Delta_{{\bf k},2},V_{{\bf k},{\bf q},12}\}\rightarrow
\{\mit\Delta_{{\bf k},2}e^{-\imath \phi},V_{{\bf k},{\bf q},12}e^{\imath
\phi}\}$ where $V_{{\bf k},{\bf q},12}=|V_{{\bf k},{\bf
q},12}|e^{\imath \phi}$, and the relative phase between the layers
would no longer be $\pm 1$.

We note that by performing the unitary transformation $\hat S^\dagger \hat
H \hat S$ where:

\begin{eqnarray*}
\hat S\equiv\frac{1}{\sqrt{2}}\left[\begin{array}{rrrr}
1&1&\ \ 0&0\\
1&-1&0&0\\
0&0&1&1\\
0&0&1&-1
\end{array}
\right],
\end{eqnarray*}

\noindent and making the substitutions $V_{11}=V_{22}=
V_{\parallel}+V_{\perp}$, $V_{12}= V_{\parallel}-V_{\perp}$,
$\varepsilon_{1}= \varepsilon+t$ and $\varepsilon_{2}= \varepsilon-t$
we obtain both the Hamiltonian and {\sc bcs} equations used by Liu
{\it et al}.\cite{liu} Our work differs from theirs in that we allow
both the dispersion and the interaction to be different in the two
layers.  This is important not only because we are able to model
systems such as YBCO in which there are CuO$_2$ planes and CuO chains,
but also because the order parameter in each of the layers may differ
in sign even in two identical layers.\cite{combescot} Other workers
have studied models in which the electrons in the pairs reside in
different layers\cite{kettemann} (ie, in which only $\chi_{{\bf
k},12}$ is non-zero) as well as models in which there is no intralayer
interaction\cite{combescot,wheatley} (ie, in which only $V_{{\bf
k,q},12}$ is non-zero).

After solving the set of coupled {\sc bcs} equations (\ref{bcs.eq}) at
$T=0$ using a {\sc fft} technique\cite{odonovan1,odonovan3,odonovan2}
we approximate the order parameters, $\mit\Delta_{{\bf k},1}$ and
$\mit\Delta_{{\bf k},2}$, with:

\begin{eqnarray}
\mit\Delta_{{\bf k},\alpha} =&&\left(\mit\Delta_\alpha^{(s_\circ)} \eta_{\bf k}^{(s_\circ)}
+\mit\Delta_\alpha^{(s_x)}\eta_{\bf
k}^{(s_x)}+\mit\Delta_\alpha^{(d)}\eta_{\bf k}^{(d)}\right) \nonumber\\
&&\times \tanh \left(1.74\sqrt{T_{\rm c}/T-1}\right),
\label{harmonics.eq}
\end{eqnarray}

\noindent
where the $\eta_{\bf k}^{(\cdot)}$ are the three lowest harmonics  given by:

\begin{eqnarray*}
\eta_{\bf k}^{(s_\circ)}&=& 1\nonumber \\
\eta_{\bf k}^{(s_x)}&=& \cos(k_x)+\cos(k_y)\nonumber \\
\eta_{\bf k}^{(d)}&=& \cos(k_x)-\cos(k_y),
\end{eqnarray*}

\noindent
and the $\mit\Delta_\alpha^{(\cdot)}$ are their amplitudes. The
amplitudes of the higher harmonics are all very much smaller in
magnitude and the gap nodes and the maximum magnitude of the gap,
which are the most important features of the order parameter, are
essentially unchanged by this approximation.  We also calculate the
magnetic penetration depth which in this system, since the
$\hat\gamma^{(i)}_{{\bf k},\alpha\beta}$ are diagonal in the greek
indices, is given by the simplified expression:

\begin{eqnarray}
\label{pd2.eq}
\lambda_{ij}^{-2}=&&\frac{4\pi e^2}{\hbar^2c^2}\frac{1}{\Omega}\sum_{{\bf k},\alpha}
\frac{\partial \varepsilon_{{\bf k},\alpha}}{\partial k_i}
\frac{\partial \varepsilon_{{\bf k},\alpha}}{\partial k_j}
\nonumber\\&&\times
\left(
\frac{\partial f(\varepsilon_{{\bf k},\alpha})}{\partial \varepsilon_{{\bf k},\alpha}}
-\frac{\partial f(E_{{\bf k},\alpha})}{\partial E_{{\bf k},\alpha}}
\right)
\end{eqnarray}

\noindent which is the usual expression\cite{odonovan2} summed
over the layers.

The curvature of the penetration depth curve, $\lambda_{ii}^{-2}(T)$,
(and also its low temperature slope) is governed by the ratio
$2\mit\Delta_{\rm max}/T_{\rm c}$, where $\mit\Delta_{\rm max}$ is the
maximum value of the order parameter in the first Brillouin zone, and
is close to a straight line for the $d$-wave {\sc bcs} value of
$2\mit\Delta_{\rm max}/T_{\rm c}=$4.4. The presence of the interlayer
interaction increases this ratio and makes the $\lambda_{ii}^{-2}(T)$
curve have a downward curvature.  Experimental measurements of both
the ratio $2\mit\Delta_{\rm max}/T_{\rm c}$\cite{edwards,maggio} as well as
the penetration depth in high quality crystals of both
YBCO\cite{basov} and BSCO\cite{jacobs} indicate that this ratio is
quite high in the {\sc htc} materials -- on the order of 6 or 7.

Other quantities calculated are the Knight shift which is given by:

\begin{equation}
K(T)\propto \frac{1}{\Omega}\sum_{\bf k} \frac{\partial f(E_{\bf k})}{\partial E_{\bf k}},
\label{ks.eq}
\end{equation}

\noindent
the normal state electronic {\sc dos} which is given by:

\begin{eqnarray}
        N(\omega)&=&\frac{1}{\Omega}\sum_{\bf k}\delta
        (\varepsilon_{\bf k}-\omega) \nonumber\\ &=&\lim_{\mit\Gamma\to
        0}\frac{1}{\pi\Omega}\sum_{\bf
        k}\frac{\mit\Gamma}{(\varepsilon_{\bf k}-\omega)^2+\mit\Gamma^2},
\label{dos.eq}
\end{eqnarray}

\noindent
the superconducting electronic {\sc dos} which is given by:

\begin{eqnarray}
        N(\omega)&=&\frac{1}{\Omega}\sum_{\bf k}\delta (E_{\bf
        k}-\omega) \nonumber\\ &=&\lim_{\mit\Gamma\to
        0}\frac{1}{\pi\Omega}\sum_{\bf k}\frac{\mit\Gamma}{(E_{\bf
        k}-\omega)^2+\mit\Gamma^2},
\label{scdos.eq}
\end{eqnarray}

\noindent
and the $c$-axis Josephson junction resistance-tunneling current
product, $RJ(T)$, through a superconductor-insulator-superconductor
junction for incoherent $c$-axis tunneling is given by the
relation~\cite{ambegaokar}:

\begin{eqnarray}
\label{joe.eq}
RJ(T)=&&\frac{2\pi T}{N^L(0) N^R(0) \pi^2}\sum_n { A^L(\omega_n)A^R(\omega_n)},
\end{eqnarray}

\noindent
where:

\begin{eqnarray*}
A^{L(R)}(\omega_n)\equiv\frac{1}{\Omega}\sum_{\bf k} \frac{\mit\Delta^{L(R)}_{\bf
k}}{(\varepsilon^{L(R)}_{\bf k})^2+(\mit\Delta^{L(R)}_{\bf k})^2+(\omega_n
)^2}.
\end{eqnarray*}

\noindent
in which the superscript L(R) indicates on which side of the junction the
dispersion and order parameter are on, the sum over $\omega_n\equiv\pi
T(2n-1)$ is for all Matsubara frequencies, $R$ is the resistance of
the junction and $N^{L(R)}(0)$ is the normal state electronic {\sc
dos} given by equation (\ref{dos.eq}). If the tunneling were coherent the
matrix element (which is incorporated into $R$) would have a $({\bf
k-k^\prime})$ dependence, and the sums over ${\bf k}$-space wouldn't be
separable.

\section{Results}

In this section we make an explicit choice for the dispersions and
interactions and then present the results of our numerical solutions
to the coupled {\sc bcs} equations (\ref{bcs.eq}) as well as the
results of our calculations of the magnetic penetration depth, densities
of states, Knight shift and Josephson current. As we wish to model
YBCO we will want to account for both the CuO$_2$ planes as well as
the CuO chains. Further, we will assume that we do not have a pairing
interaction in the chains, but only in the planes as well as an
interlayer interaction. This means that all of the order parameter in
the chains is due to the interlayer interaction.
We note that although our solution technique\cite{odonovan1} allows
the order parameters to be complex and to have a relative phase
between layers we find that in the models studied here, to within an
overall phase, the order parameters are all real with a relative phase
of $\pm 1$.

\begin{figure}[t]
\postscript{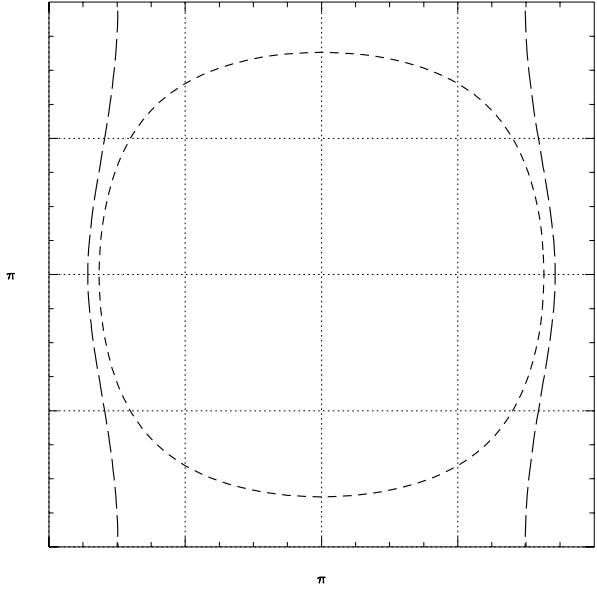}
\caption{
Model of YBCO Fermi surfaces for chains (long dashed curve) and planes
(closed short dashed curve) in the first Brillouin zone. The
$(\pi,\pi)$ point is at the center of the figure.  For the chains the
parameters $\{t_\alpha,\epsilon_\alpha,B_\alpha,\mu_\alpha\}$ in
Eq.~\protect\ref{disp.eq} are $\{-50,-0.9,0,1.2\}$ and for the planes
they are $\{100,0,0.45,0.51\}$.  }
\label{fermi.fig}
\end{figure}

For the dispersions, $\varepsilon_{{\bf k},\alpha}$, we use:

\begin{eqnarray}
\label{disp.eq}
\varepsilon_{{\bf k},\alpha}=
&-&2t_\alpha\left[(1+\epsilon_\alpha)\cos(k_x)+\cos(k_y)\right.\nonumber\\
&-&\left.2B_\alpha\cos(k_x)\cos(k_y)-(2-2B_\alpha-\mu_\alpha)\right],
\end{eqnarray}

\noindent where the parameters $\{t_\alpha, \epsilon_\alpha, B_\alpha,
\mu_\alpha\}$ are chosen so that the Fermi surface and bandwidth are
close approximations to those observed experimentally \cite{gofron}
(see Fig.~\ref{fermi.fig}). In order to model YBCO we chose
$\{100,0,0.45,0.51\}$ for the planes and $\{-50,-0.9,0,1.2\}$ for the
chains. In both dispersions $t_\alpha$, which sets the overall energy
scale, is in units of meV.  For the interactions, $V_{{\bf k},{\bf
q},\alpha\beta}$, we chose an {\sc mmp} form:\cite{mmp}

\begin{eqnarray*}
V_{{\bf k},{\bf q},\alpha\beta}=g_{\alpha\beta}
\frac{-\chi_\circ}{1+\xi_\circ^2|{\bf k-q-Q}|^2},
\end{eqnarray*}

\noindent where $\chi_\circ$ is a constant that sets the scale of the
susceptibility, $\xi_\circ$ is the magnetic coherence length, ${\bf
Q}\equiv(\pi,\pi)$ is the commensurate nesting vector, and
$g_{\alpha\beta}$ is the coupling to the conduction electrons, the
size of which can be fixed to get a desired value of the critical
temperature and can be considered to contain $\chi_\circ$. The
remaining parameter, $\xi_\circ$, is given in reference \cite{mmp} and
will not be varied in this work.  In this paper we set $g_{22}=0$ ie,
there is no intrinsic pairing in the chains. This means that any
superconductivity in the chains is induced by the interlayer
interaction, $g_{12}$, since we have set the hopping between layers to
zero. The effect of an interlayer hopping has been extensively studied
in works by Atkinson and Carbotte\cite{atkinson} as well as
others.\cite{klemm,klemm2,kettemann,wheatley,liu}

\begin{figure}[t]
	\begin{center} \begin{tabular}{c c} 
		\makebox[0.5\columnwidth][l]{\large (a)} &
		\makebox[0.5\columnwidth][l]{\large (b)} \\
		\epsfxsize=0.5\columnwidth \epsfbox{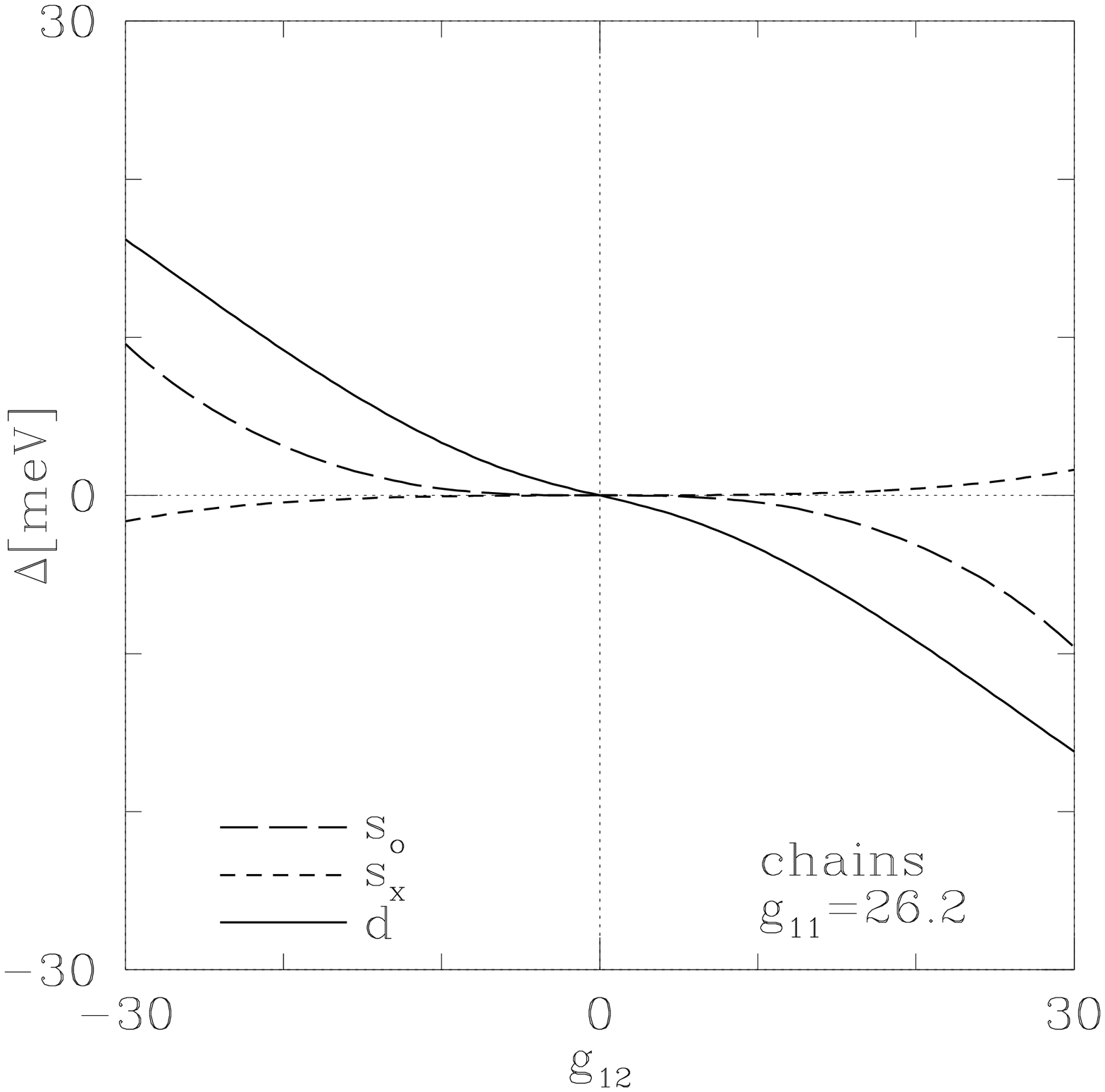} &
		\epsfxsize=0.5\columnwidth \epsfbox{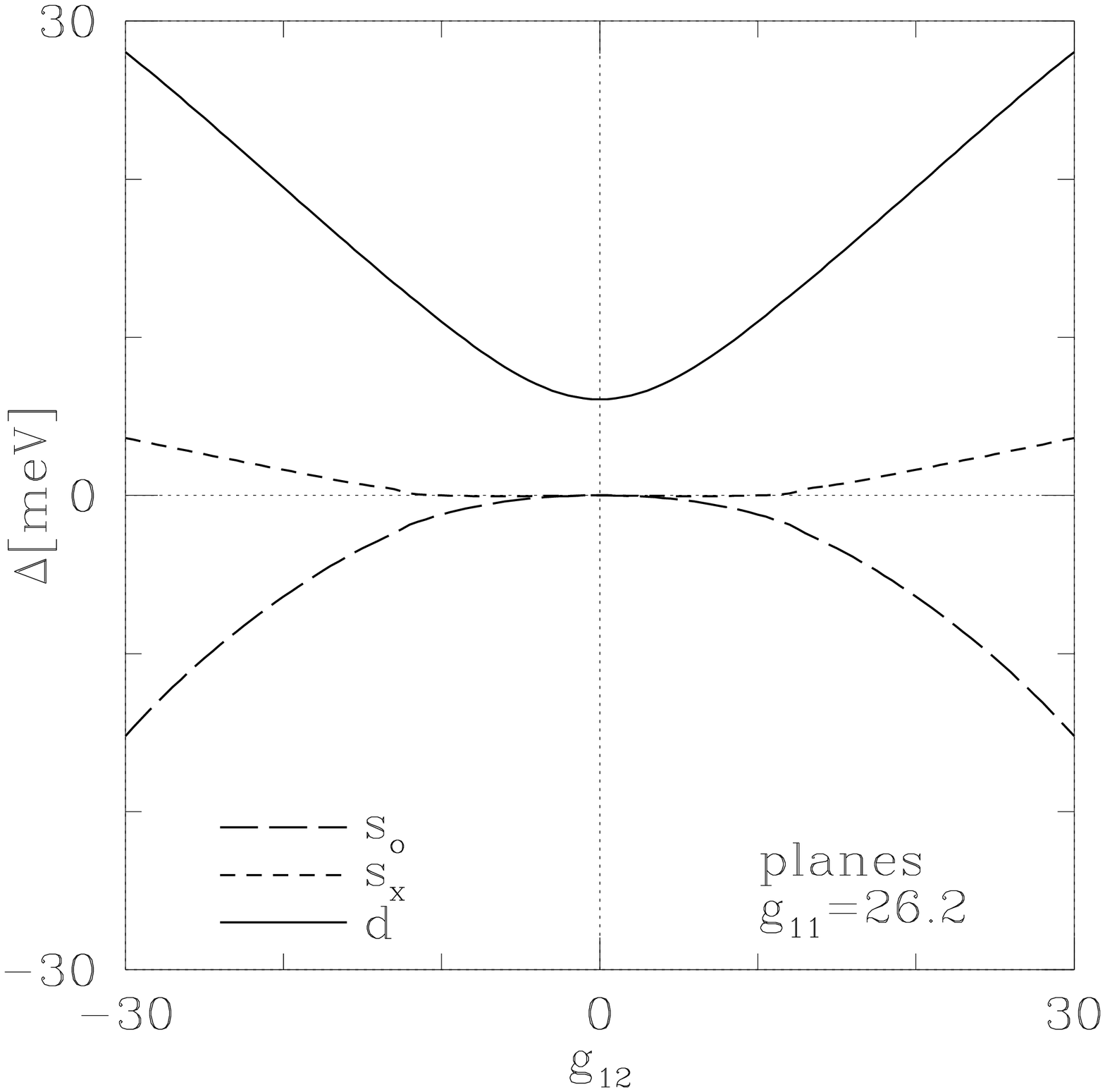} \\
		\makebox[0.5\columnwidth][l]{\large (d)} \\
		\epsfxsize=0.5\columnwidth \epsfbox{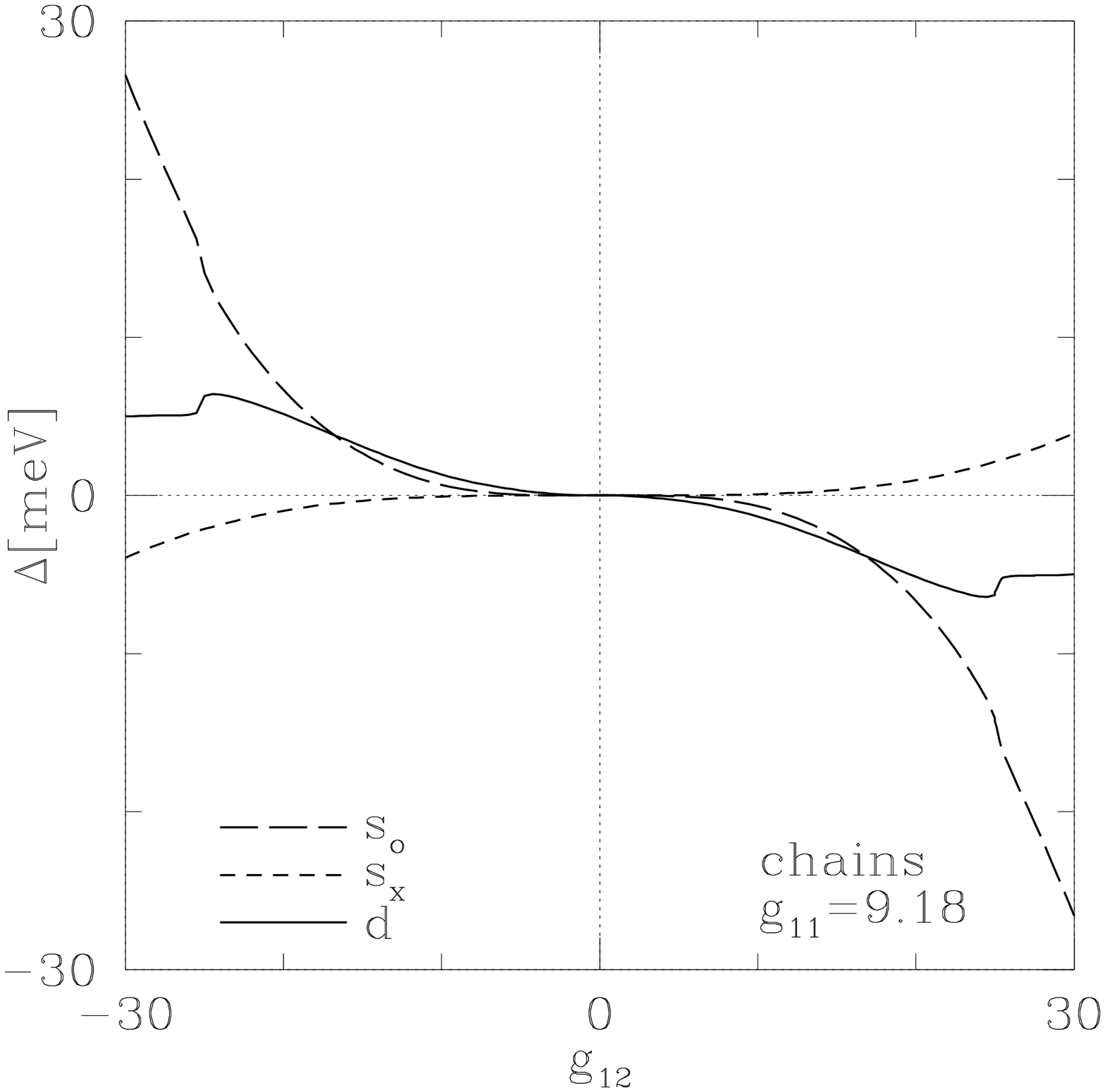} &
		\epsfxsize=0.5\columnwidth \epsfbox{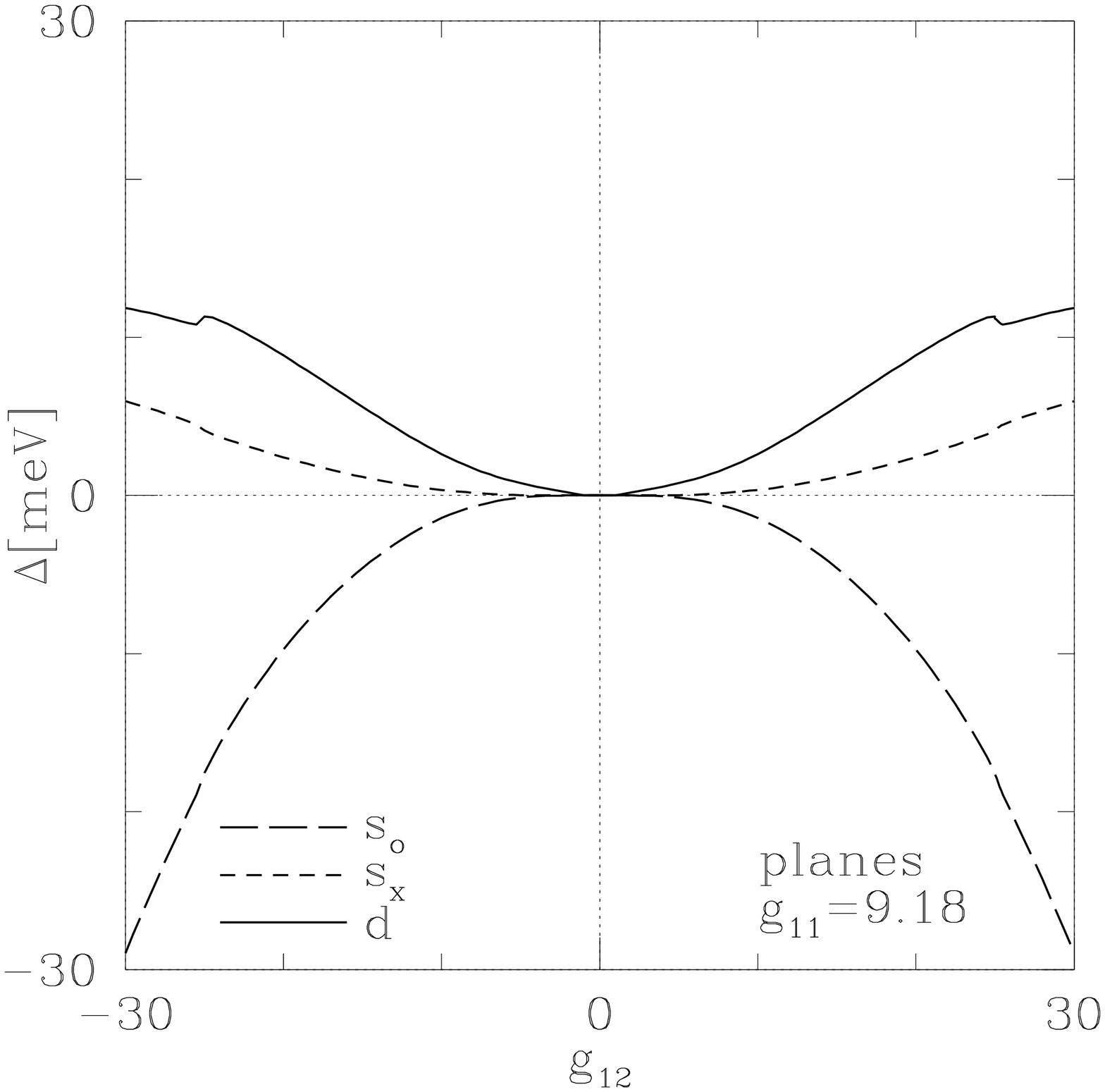} \\
	\end{tabular} \end{center}
\caption{ 
Calculation of the zero temperature order parameters as a function of
the interlayer interaction, $g_{12}$, for two fixed values of the
interlayer interaction, $g_{11}$, (upper and lower frames) presented
for the planes (right frames) and chains (left frames) separately. In
all frames the solid curve is the $d$-wave component of the order
parameter, the short dashed curve is the extended $s$-wave component
and the long dashed curve is the isotropic $s$-wave component. In the
upper frames $g_{11}=26.2$ and for $g_{12}=10$, $T_{\rm c}=100K$. At
$g_{12}=0$ the order parameter is zero in the chains and is pure
$d$-wave in the planes. As the interlayer interaction is increased the
order parameter becomes present in the chains and there is a mixing of
$s$-wave components. In the lower frames $g_{11}=9.18$ and for
$g_{12}=20$, $T_{\rm c}=100K$. At $g_{12}=0$ the order parameter is zero in
both the chains and the planes. As the interlayer interaction is
increased the order parameter becomes present in both the chains and
planes and there is a mixing of $s$-wave components with the isotropic
$s$-wave component eventually becoming dominant. The feature at
$g_{12}\sim 25$ occurs when the gap node leaves the Brillouin zone.
As discussed in the text there is a $g_{12}\leftrightarrow -g_{12}$
symmetry. Both $g_{11}$ and $g_{12}$ are in units of $t_1$; $g_{22}$,
the coupling in the chains, is set equal to zero.}
\label{ybco.fig}
\end{figure}

We solve the coupled {\sc bcs} equations (\ref{bcs.eq}) using a {\sc
fft} technique.\cite{odonovan1} In Fig.~\ref{ybco.fig} we plot the
lowest three Fourier components (\ref{harmonics.eq}) of the zero
temperature order parameter as a function of the interlayer
interaction (higher Fourier components are all much smaller in
magnitude) for two different intralayer interactions (upper and lower
frames) in the planes (left frames) and chains (right frames). The
values plotted in Fig.~\ref{ybco.fig} are the amplitudes
$\mit\Delta_\alpha^{(s_\circ)}$, $\mit\Delta_\alpha^{(s_x)}$ and
$\mit\Delta_\alpha^{(d)}$ given by (\ref{harmonics.eq}) with
$\alpha=1$ for the plane layers (right frames) and $\alpha=2$ for the
chain layers (left frames). For the orthorhombic system studied here
all three of these harmonics belong to the same irreducible
representation of the crystal point group except for $g_{12}=0$ when
the tetragonal CuO$_2$ layer is decoupled from the orthorhombic CuO
layer. Recent {\sc cits} measurements\cite{edwards2} show that the gap
has a magnitude of $\sim 20$meV in the chains and $\sim 30$meV in the
planes which would indicate that $g_{12}$ is large.

For the first choice of intralayer interaction (upper frames),
$g_{11}=26.2$, and there is no order parameter in the chains when
there is no interlayer interaction (ie, $g_{12}=0$) and the order
parameter in the planes is pure $d$-wave. As the interlayer
interaction is increased from zero, $s$-wave components appear in the
planes and all three components appear in the chains. This ``$s+d$
mixing'' is caused by the breaking of the tetragonal symmetry upon the
introduction of the chains; there is no relative phase between the
$s$- and $d$-wave components within either the planes or chains but
there can be a relative phase between the order parameter in the
planes and chains. In the range of $g_{12}$ explored here the $d$-wave
component in the plane remains dominant but for sufficiently strong
interlayer interaction the isotropic $s$-wave component eventually
dominates\cite{liechtenstein} (ie, the gap nodes disappear). 
For interaction parameters $\{g_{11},g_{12}, g_{22}\} = \{26.2,10,0\}$
the critical temperature is 100K and the maximum value of the gap in
the Brillouin zone is 27.5meV in the planes and 8.0meV in the chains,
while the maximum values on the Fermi surfaces are approximately 22meV
and 7meV respectively. The ratio $2\mit\Delta_{\rm max}/T_{\rm c}$ is 6.4 in the
planes and 1.9 in the chains.

For the second choice of intralayer interaction (lower frames),
$g_{11}=9.18$, there is no order parameter in either the chains or the
planes when there is no interlayer interaction (ie, $g_{12}=0$). As
the interlayer interaction is increased $d$-wave and then $s$-wave
components of the order parameter appear in both the planes and
chains. Again, there is no relative phase between the $s$- and
$d$-wave components within either the planes or chains but there can
be a relative phase between the order parameter in the planes and
chains. In the range of $g_{12}$ explored here, the $d$-wave component
is dominant. At approximately $g_{12}=15$ the gap nodes no longer
cross the Fermi surface in the chains; the feature at $g_{12}\sim 25$
coincides with the gap nodes leaving the Brillouin zone and isotropic
$s$-wave becoming dominant.  For interaction parameters
$\{g_{11},g_{12}, g_{22}\} = \{9.18,20,0\}$ the critical temperature
is again 100K and the maximum value of the gap in the Brillouin zone
is now 32.8meV in the planes and 20.1meV in the chains, while the
maximum values on the Fermi surfaces are approximately 27meV and 17meV
respectively. The ratio $2\mit\Delta_{\rm max}/T_{\rm c}$ is 7.6 in the
planes and 4.7 in the chains.

Note that for $g_{12}>0$  all of the $s$-wave components of the
order parameters in both the planes and chains have the same relative
sign and the $d$-wave components have opposite signs, while for
$g_{12}<0$ all of the relative signs are reversed but that the
magnitudes of the components are insensitive to the sign of
$g_{12}$ as noted after Eq.~\ref{bcs.eq}.

\begin{figure}[t]
	\begin{center}
	\begin{tabular}{c c} 
		\makebox[0.5\columnwidth][l]{\large (a)} &
			\makebox[0.5\columnwidth][l]{\large (b)} \\
			\epsfxsize=0.5\columnwidth
			\epsfbox{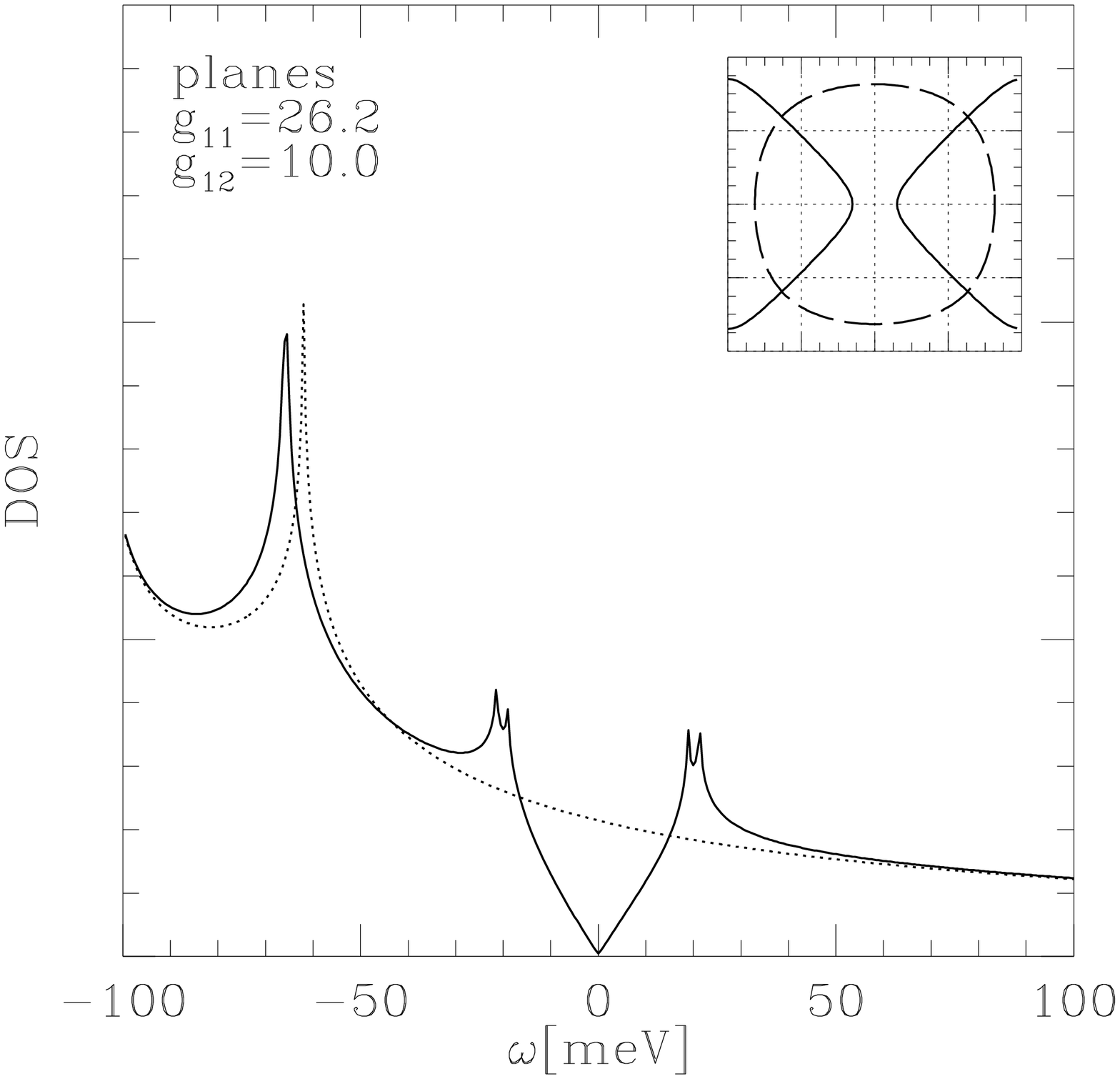} &
			\epsfxsize=0.5\columnwidth
			\epsfbox{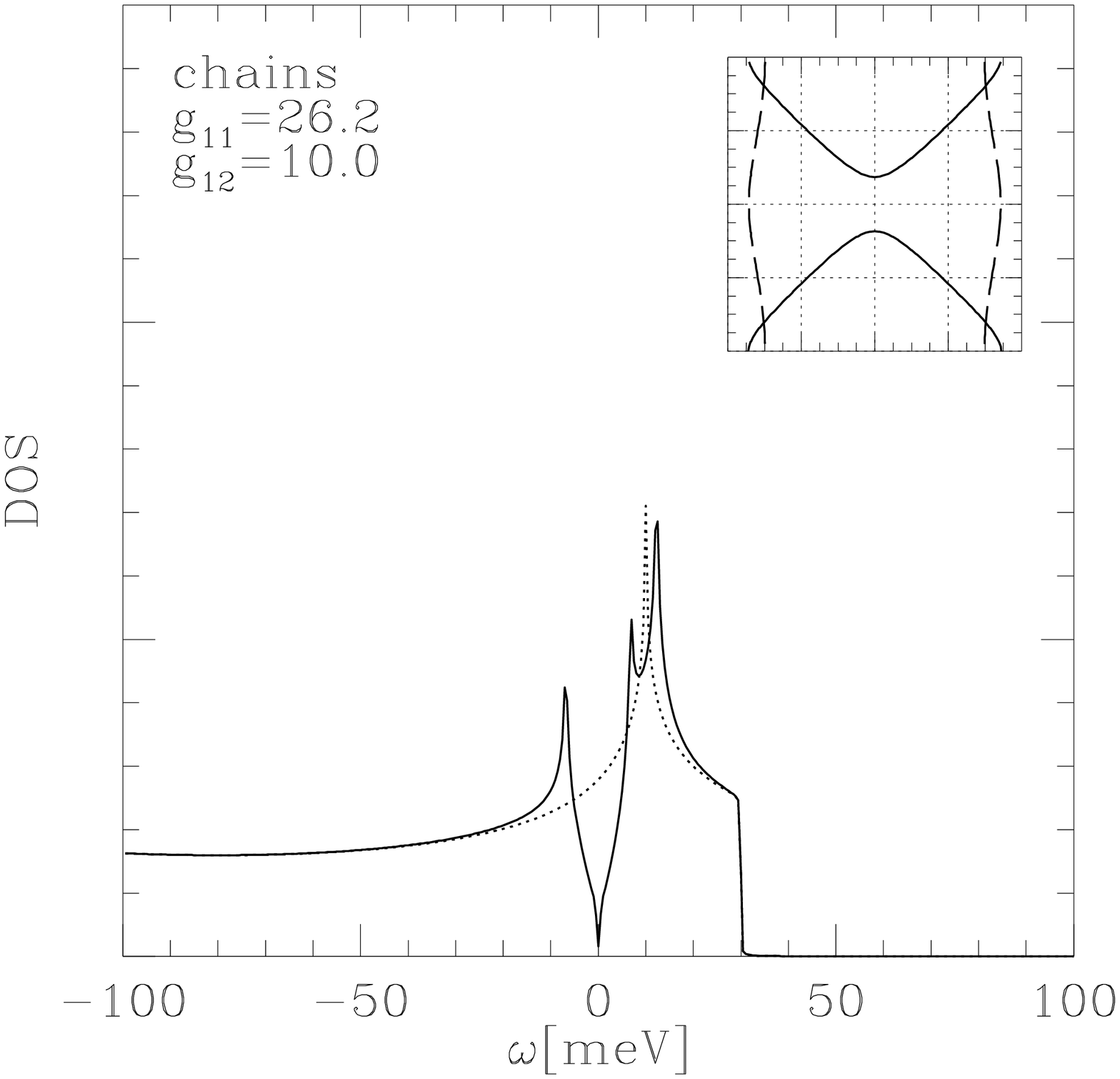} \\ 
			\makebox[0.5\columnwidth][l]{\large (c)} &
			\makebox[0.5\columnwidth][l]{\large (d)} \\
			\epsfxsize=0.5\columnwidth
			\epsfbox{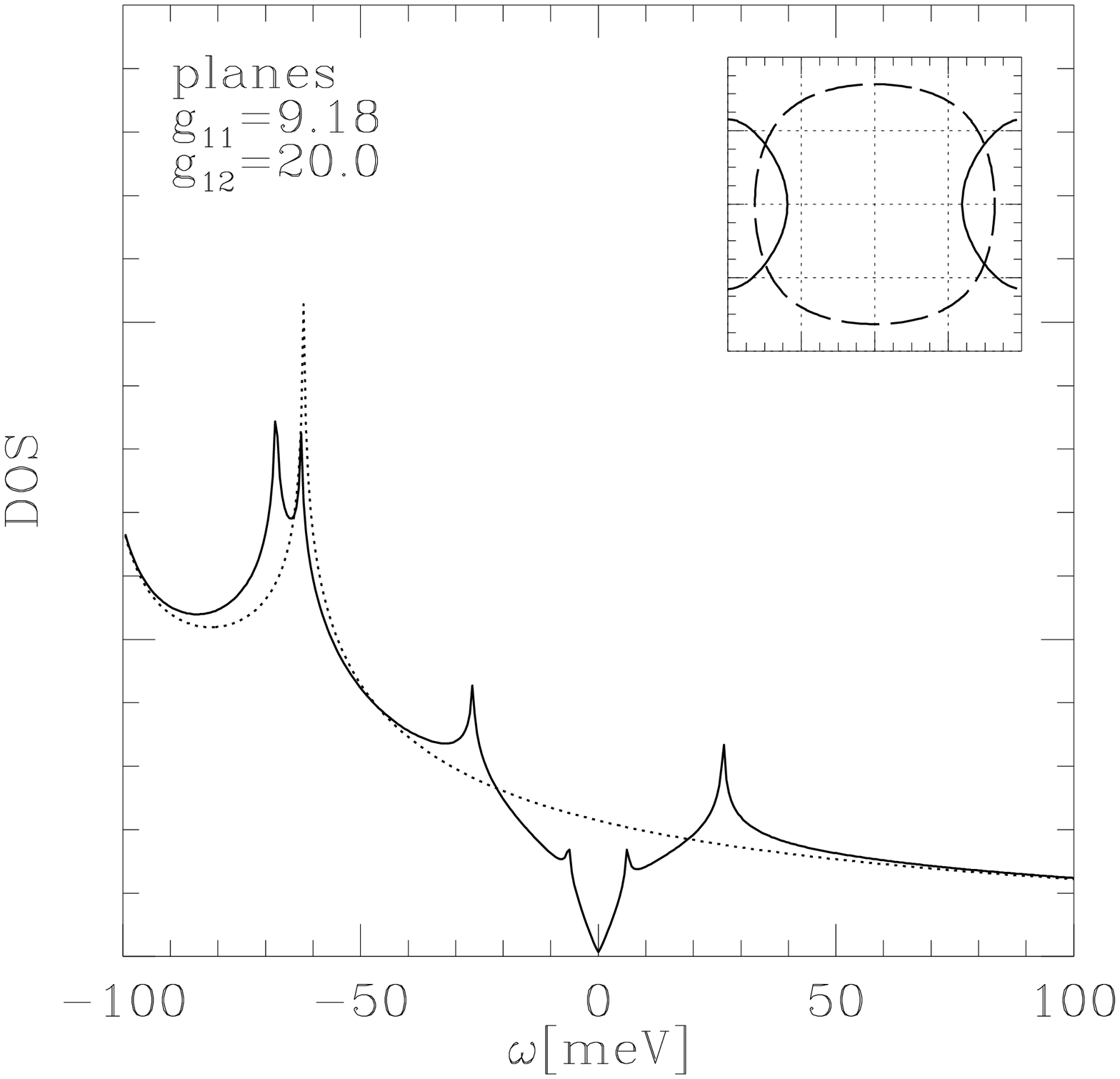} &
			\epsfxsize=0.5\columnwidth
			\epsfbox{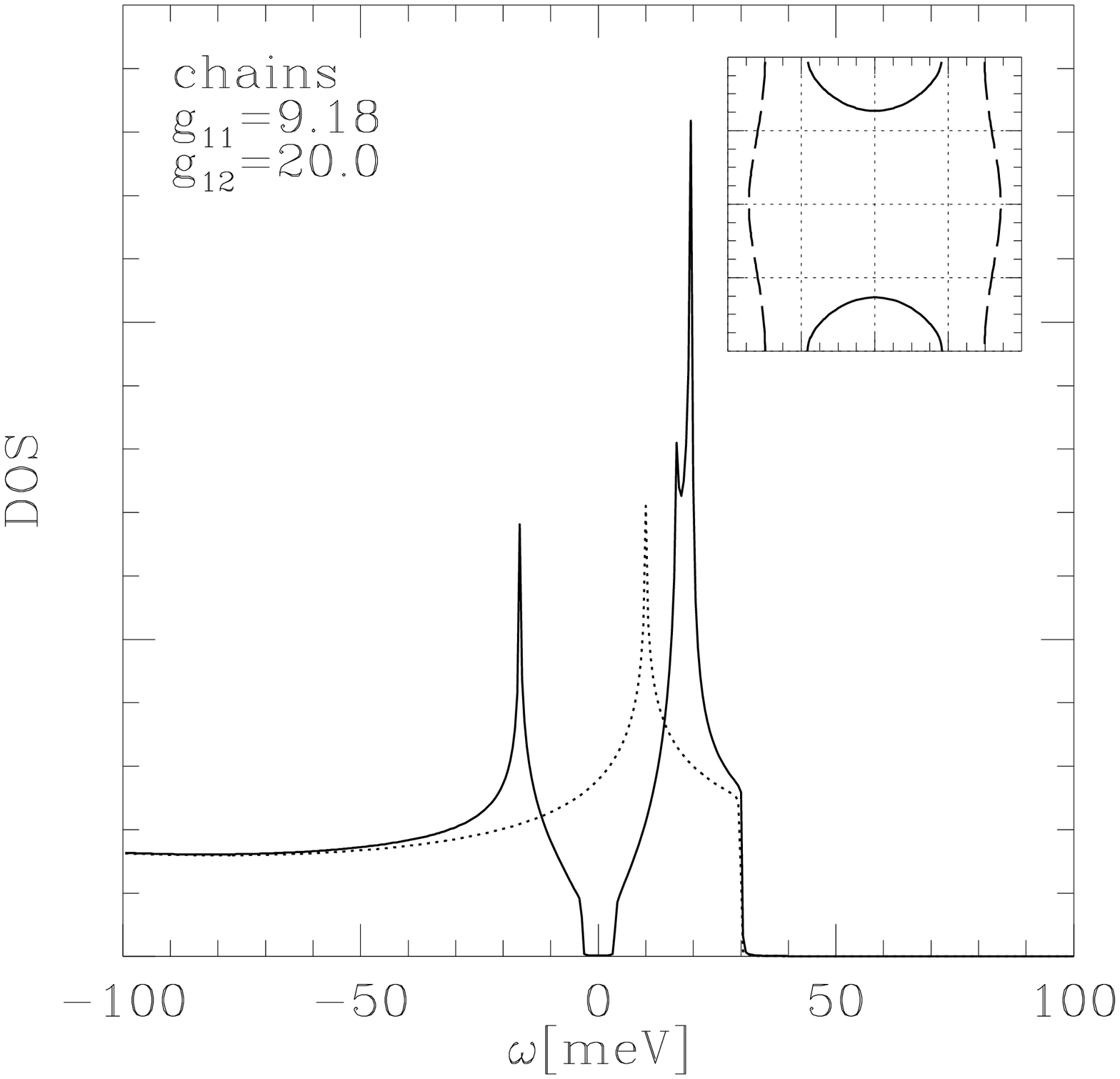} \\ 
		\end{tabular}
	\end{center}
\caption{ Calculation of the superconducting (solid curves) and normal
(dotted curves) density of states ({\sc dos}) for two sets of
interaction parameters $g_{\alpha\beta}$ (upper and lower frames)
presented separately for the planes (left frames) and chains (right
frames). Some experiments are surface probes and may probe either the
planes or chains independently. The insets show the Fermi surface
(dashed curve) and gap nodes (solid curve) in the planes and chains
for the two different parameter choices in the first Brillouin zone
with $(\pi,\pi)$ at the center. In the upper frames, (a) and (b), we
have chosen $\{g_{11},g_{12},g_{22}\} = \{26.2,10,0\}$ which gives
$T_{\rm c}=100K$.  In the lower frames, (c) and (d), we have chosen
$\{g_{11},g_{12},g_{22}\} =\{9.18,20,0\}$ which also gives
$T_{\rm c}=100K$. Note that for the second parameter choice the gap nodes do
not cross the Fermi surface in the chains (frame (d), inset) and that
the {\sc dos} is gapped. The $c$-axis Josephson resistance-tunneling
current product for a YBCO-Pb junction for a pure $d$-wave order
parameter is zero due to the equal parts of the order parameter with
opposite signs. Here this is not the case (insets) and the $c$-axis
Josephson resistance-tunneling currents product for a YBCO-Pb junction
are (a) $\pm$0.18meV, (b) 2.35meV, (c) $\pm$0.25meV, and (d)
2.17meV. The relative sign is due to the relative sign of the $s$-wave
components of the order parameter (ie,~the only part which
contributes). }
\label{dos.fig}
\end{figure}

In Fig.~\ref{dos.fig} we plot the density of states ({\sc dos}) for
the planes (left frames) and chains (right frames) calculated using
the lowest three harmonics (\ref{harmonics.eq}) of the solution to the
{\sc bcs} equations (\ref{bcs.eq}) with two sets of interaction
parameters.  The dotted curves are the normal state {\sc dos} ({\sc
nsdos}) and the solid curves are the superconducting {\sc dos} ({\sc
scdos}).  The insets show the Fermi surface (dashed curves) and gap
nodes (solid curves) in the first Brillouin zone (with $(\pi,\pi)$ at
the center). The peak in the {\sc nsdos} (dotted curves) is the van
Hove singularity located at
$2t_\alpha(2-\mu_\alpha-4B_\alpha\pm\epsilon_\alpha)$ which is at
-62meV in the plane layers (left frames) and at 10 and -170meV in the
chain layers (right frames). They are caused by the saddle points in
the electron dispersions, $\varepsilon_{{\bf k},\alpha}$, at
$(0,\pm\pi)$ and $(\pm\pi,0)$. These van Hove singularities are
shifted by the presence of the superconducting order parameter (solid
curves) by an amount that depends upon the value of the order
parameter at the saddle points;\cite{zhou} in frame (c) these values
are very different and the van Hove singularity is split, in frame (a)
these values are almost the same and no splitting is evident.

\begin{figure}[t]
\postscript{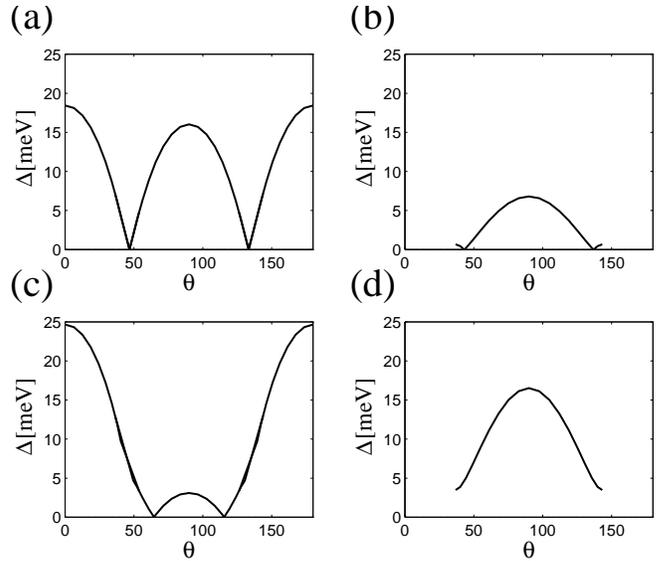}
\caption{
The magnitude of the gap on the Fermi surface as a function of angle
for the four cases of Figs.~\ref{ybco.fig} and \ref{dos.fig}. The angle
$\theta$ is measured from the center or $(\pi,\pi)$ point of the
Brillouin zone with the $y$-axis (ie, the vertical in the insets of
Fig.~\ref{dos.fig}) corresponding to $\theta=0$. Frames (b) and (d) do
not span all angles due to the Fermi surface not being closed in the
chain layer. For the first choice of interaction parameters,
$\{g_{11},g_{12},g_{22}\} = \{26.2,10,0\}$, the ratio
$2\mit\Delta_{\rm max(FS)}/T_{\rm c}$, where $\mit\Delta_{\rm max(FS)}$ is
the maximum value of the gap on the Fermi surface, is 4.3 and 1.6 for
the planes and chain respectively; for the second,
$\{g_{11},g_{12},g_{22}\} =\{9.18,20,0\}$, they are 5.7 and 3.8.}
\label{angles.fig}
\end{figure}

An interesting feature is that the low energy behaviour ($\omega\sim
0$) of the {\sc scdos} is governed by the {\em smallest} local {\em
maxima} of the gap on the Fermi surface when the gap nodes cross the
Fermi surface and by the {\em minima} of the gap on the Fermi surface
when there is no gap nodes which cross the Fermi surface.  In
Fig.~\ref{angles.fig} the magnitude of the gap along the Fermi surface
is plotted.  In Fig.~\ref{angles.fig}a-c one can see that there are
two different local maxima of the gap on the Fermi surface and these
maxima are (to first order\cite{zhou}) manifested as twin peaks in the
{\sc scdos} (Fig.~\ref{dos.fig}a-c); these peaks are distinct from the
van Hove singularities which are also present in the normal {\sc dos}
(dotted curves in Fig.~\ref{dos.fig}) and which are slightly shifted
in the superconducting state.\cite{zhou} In frame (a) of
Fig.~\ref{angles.fig} the local maxima of the gap on the Fermi surface
are 16 and 18meV; in (b) they are 1 and 7meV, and in (c) they are 25
and 3meV. In (d) one can see that there are no gap nodes which cross
the Fermi surface; the maximum and minimum value of the gap on the
Fermi surface are 17 and 4meV respectively. In Fig.~\ref{dos.fig}d the
finite gap in the {\sc scdos} corresponds to the minimum of the gap
on the Fermi surface and the peak to the maximum.

The Josephson junction resistance-tunneling current product,
$RJ(T=0)$, for a $c$-axis YBCO-Pb junction, given by (\ref{joe.eq})
with $\mit\Delta^L_{\bf k}$ and $\varepsilon^L_{\bf k}$ appropriate
for Pb,\cite{odonovan3} are $\pm0.25$meV and $\pm2.2$meV for the
planes and 2.2meV and 3.4meV for the chains for the two choices of
$g_{\alpha\beta}$ made, in agreement with earlier
calculations.\cite{odonovan3} The relative sign is due to the relative
sign of the $s$-wave components (ie, the only part which contributes)
of the order parameters. The actual $c$-axis Josephson junction
resistance-tunneling current product, $RJ(T)$, for a junction made
with untwinned YBCO would be some weighted average of the plane and
chain $c$-axis tunneling currents depending upon the relative
abundance of chains and planes in the top layer of the YBCO. For a
twinned sample with both twins equally abundant there would be zero
net tunneling current, although there is evidence that for single
crystals of YBCO there can be up to a 5:1 ratio in the relative
abundance of the two twin orientations.\cite{gaulin} We note that due
to the different magnitudes of the order parameters in the two layers
the model presented here is consistent with the observed $\pi$ shifts
in corner junctions\cite{wollman,brawner,tsuei,mathai} for both
attractive and repulsive interlayer interactions, $g_{12}$.

\begin{figure}[t]
	\begin{center}
		\begin{tabular}{c c} 
			\makebox[0.5\columnwidth][l]{\large (a)} &
			\makebox[0.5\columnwidth][l]{\large (b)} \\
			\epsfxsize=0.5\columnwidth
			\epsfbox{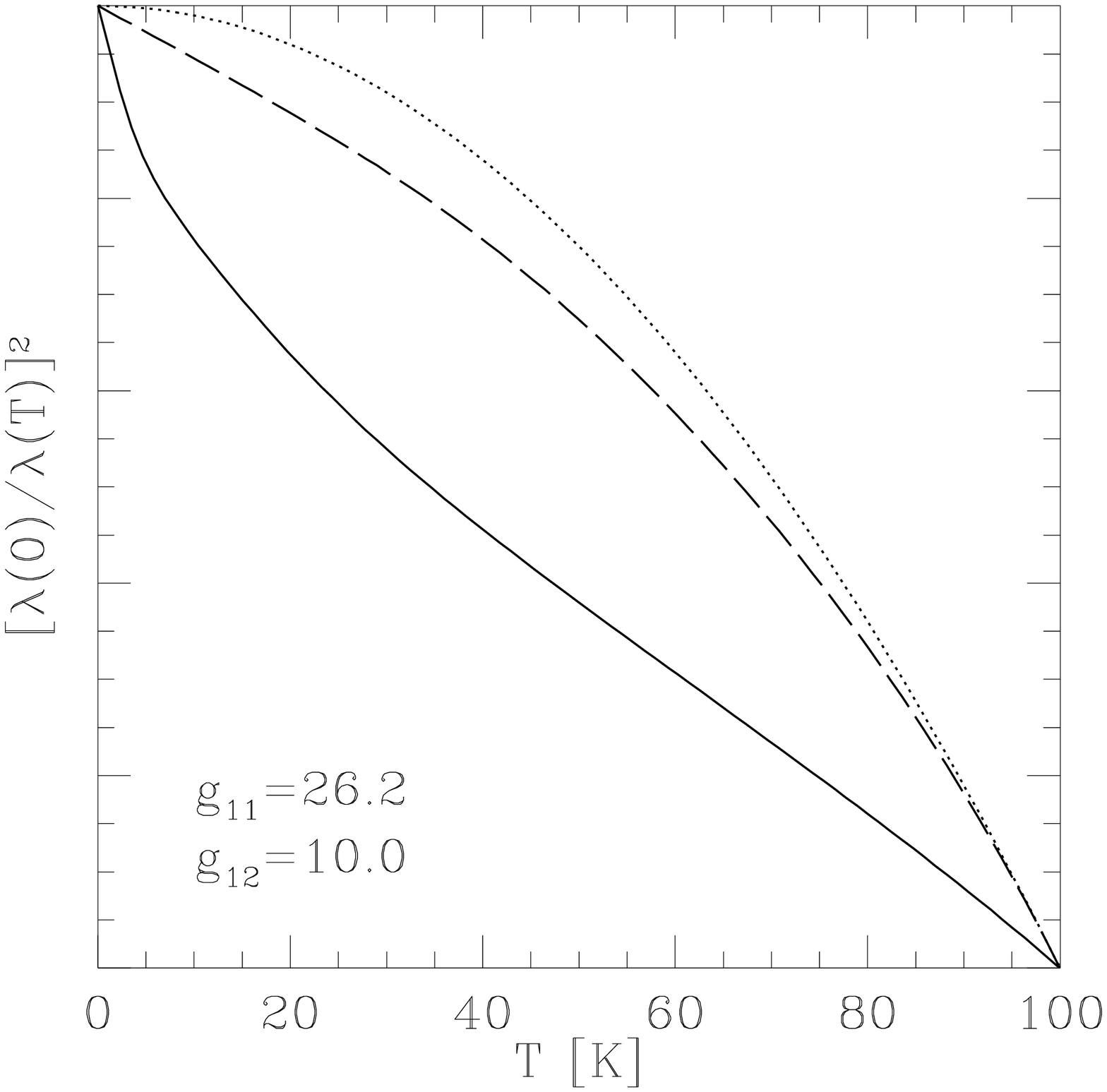} &
			\epsfxsize=0.5\columnwidth
			\epsfbox{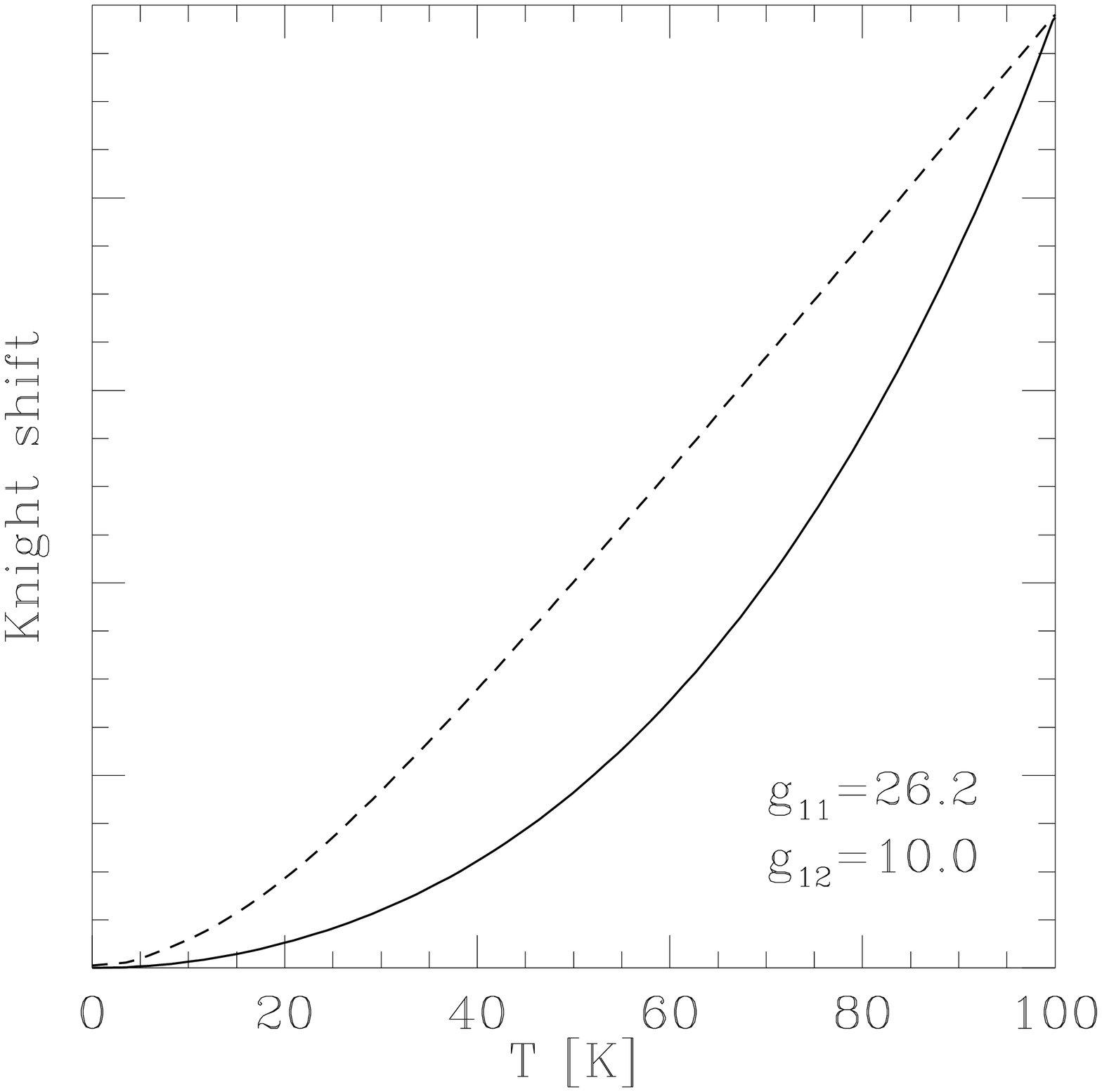} \\ 
			\makebox[0.5\columnwidth][l]{\large (c)} &
			\makebox[0.5\columnwidth][l]{\large (d)} \\
			\epsfxsize=0.5\columnwidth
			\epsfbox{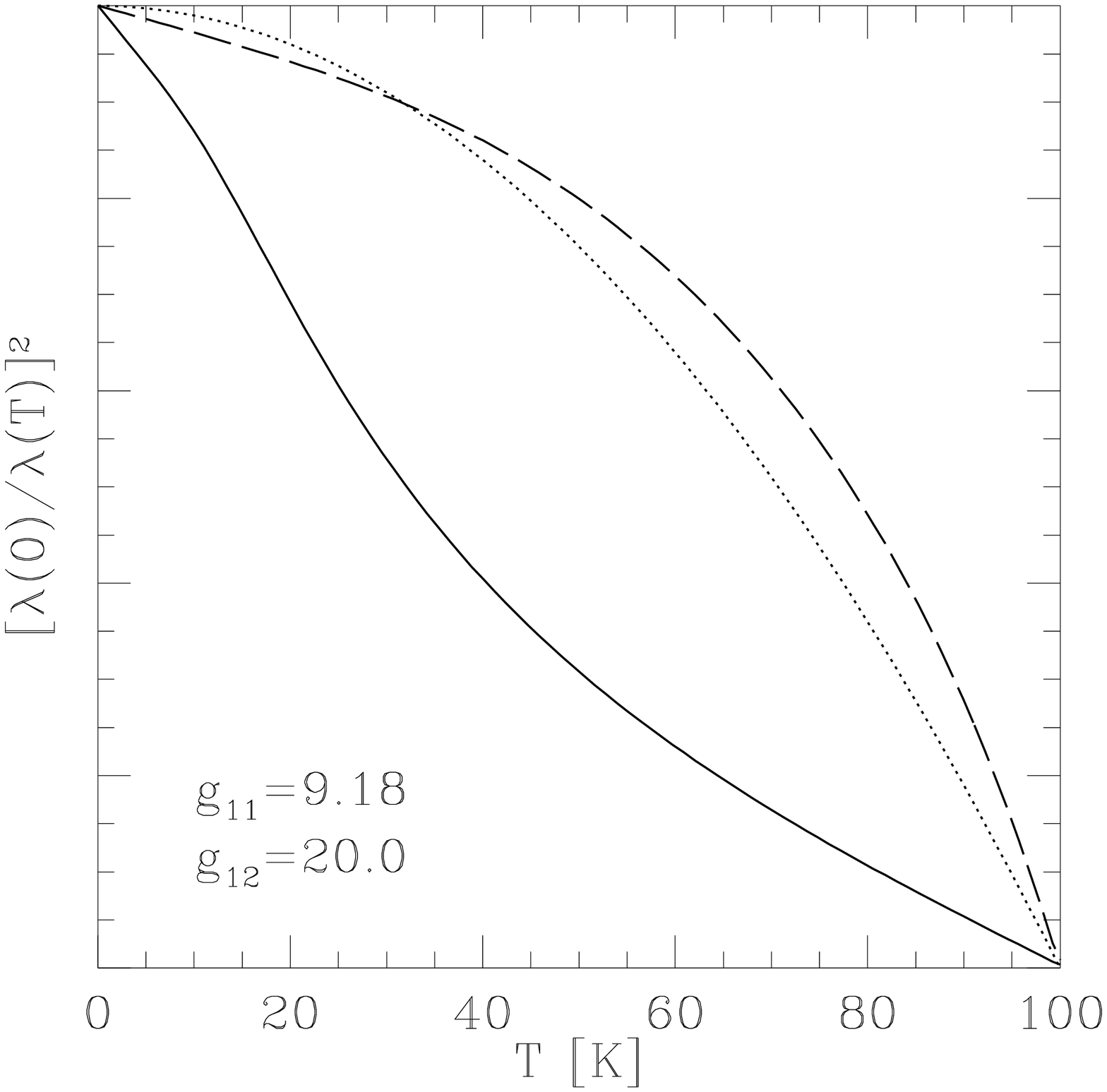} &
			\epsfxsize=0.5\columnwidth
			\epsfbox{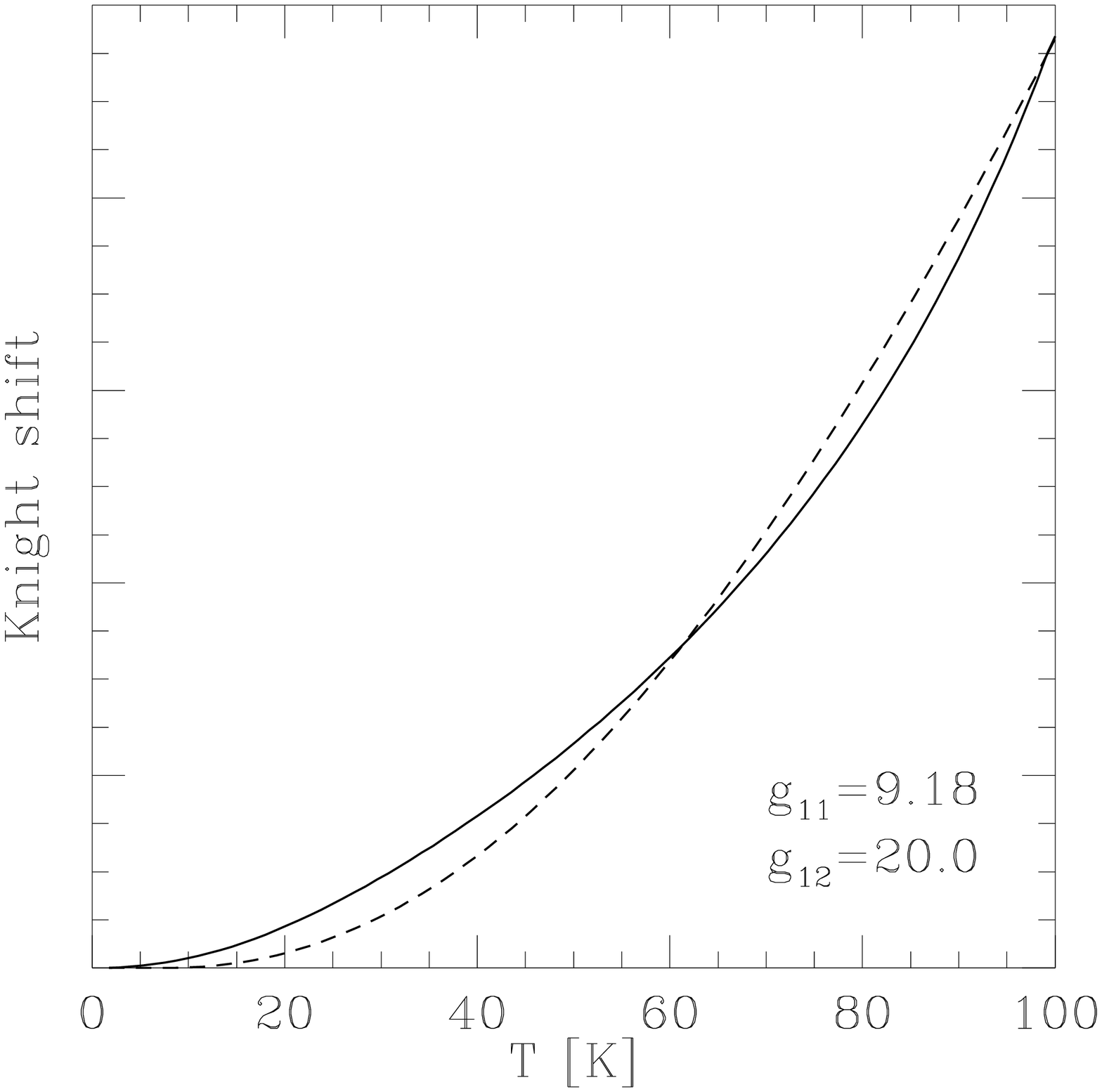} \\ 
		\end{tabular}
	\end{center}
\caption{ Calculations of the magnetic penetration depth (left frames)
and the Knight shift (right frames) for the two sets of interaction
parameters.  Frames (a) and (c) show the magnetic penetration depth
along (solid curve) and perpendicular (dashed curve) to the CuO
chains.  The dotted curve is $1-(T/T_{\rm c})^2$ and is shown for
comparison. The chains, due to their Fermi surface, do not contribute
appreciably to the penetration depth perpendicular to the chains
(dashed curves).  The ratio $(\lambda_{yy}/\lambda_{xx})^2$ is 1.37 for
both sets of parameters since this is a normal state property. Frames
(b) and (d) show the Knight shift in the planes (solid curves) and
chains (dashed curves); due to the crossing of the gap nodes and the
Fermi surface in the chains in (b) the Knight shift in the chain is a
power law at low temperature, while due to the finite gap in (d) the
Knight shift in the chain is exponential at low temperature.}
\label{pd.fig}
\end{figure}

In Fig.~\ref{pd.fig} we have plotted the magnetic penetration depth
(left frames) and the Knight shift (right frames) calculated with the
lowest three harmonics (\ref{harmonics.eq}) of the solutions of the
{\sc bcs} equations (\ref{bcs.eq}) for the two choices of interaction
parameters.  In the penetration depth frames (left) the solid curve is
for the $x$-direction (along the chains) and the dashed curve is for
the $y$-direction (perpendicular to the chains). The dotted curve is
$1-(T/T_{\rm c})^2$ and is plotted for comparison. The ratio
$\lambda_{yy}/\lambda_{xx}$ at zero temperature is 1.37 for both
interaction parameter choices since the zero temperature penetration
depth is a normal state property (ie, the second term in
Eq.~\ref{pd2.eq} does not contribute at zero temperature). The zero
temperature penetration depth is largely governed by the bandwidth
(ie, $4t_\alpha(2-\epsilon_\alpha)$) -- the larger the bandwidth the
larger the zero temperature penetration depth.

As pointed out above, the curvature of the penetration depth curve,
$\lambda_{ii}^{-2}(T)$, is largely governed by the ratio $2\mit\Delta_{\rm
max}/T_{\rm c}$ and is a straight line for the $d$-wave {\sc bcs} value of
$4.4$. The presence of the chain layer and the interlayer interaction
increases this ratio in the plane layer but it remains low in the
chain layer due to the absence of an interaction in this layer. It is
this lower value that makes $\lambda^{-2}_{yy}(T)$ (along the chains)
have upward curvature (solid curves).

One can see (Fig.~\ref{pd.fig}a,c) that the in-plane penetration depth
perpendicular to the chains (dashed curve) closely resembles that
observed experimentally in high quality
crystals,\cite{basov,jacobs} and is largely determined by the
presense of gap nodes crossing the Fermi surface, which cause the low
temperature linear behaviour, and the ratio $2\mit\Delta_{\rm
max}/T_{\rm c}$ which, for values above $\sim4.4$, make the curve of
$\lambda_{ii}^{-2}(T)$ have downward curvature. The penetration depth
along the chains (solid curves), however, has an overall upward
curvature due to the low values of $2\mit\Delta_{\rm max}/T_{\rm c}$ in the
chains. It is the component of the penetration depth due to the chains
that makes the overall $\lambda^{-2}_{yy}(T)$ have downward
curvature. The component of the penetration depth due to the chains
perpendicular to the chains does not contribute significantly to the
overall penetration depth in this direction, $\lambda^{-2}_{yy}$, is due
almost entirely to the CuO$_2$ layers. For a single plane
model\cite{odonovan2} we would have $\{g_{11},g_{12}, g_{22}\}
=\{30,0,0\}$, $2\mit\Delta_{\rm max}/T_{\rm c}$ would be 4.5, and the
penetration depth would closely resemble the straight line
$1-(T/T_{\rm c})$.\cite{odonovan2}

The Knight shift (Fig.~\ref{pd.fig}b,d), which is calculated
independently for the planes (solid curves) and chains (dashed
curves), has a low temperature power law behaviour when the gap nodes
cross the Fermi surface (planes, both figures and chains in upper
figure) and an exponential behaviour when the gap is finite over all
the Fermi surface (chains, lower figure). When these quantities are
measured\cite{scalapino} the distinction between a power law and
exponential behaviour rests upon the choice zero and so is not a
reliable indicator of the presence of gap nodes on the Fermi surface.

\section{Conclusion and Discussion}

We have derived a general expression for the Hamiltonian in a
multi-layer system and then made a simplification and have explicitly
diagonalized the Hamiltonian. A set of two coupled {\sc bcs} equations
are then derived for this simplified system which is subsequently
solved numerically by a {\sc fft} technique. This technique, unlike
others,\cite{wheatley,combescot} makes no assumptions about the
functional form (and hence the symmetry) of the order parameter in
either layer nor any relationship between the order parameters in the
different layers.

Using the three lowest harmonics (\ref{harmonics.eq}) of the solutions
found for the coupled {\sc bcs} equations (\ref{bcs.eq}) the magnetic
penetration depth (\ref{pd2.eq}), normal and superconducting density
of states (\ref{dos.eq},\ref{scdos.eq}), Knight shift (\ref{ks.eq})
and $c$-axis Josephson resistance-tunneling current products
(\ref{joe.eq}) were calculated.

Our choice of electron dispersion relations were made so as to
approximate the YBCO system in which there are layers consisting of
CuO$_2$ planes as well as layers which contain CuO chains. Our choice
of interactions was made so that there is a pairing interaction in the
CuO$_2$ layer as well as an interlayer interaction, but no pairing
interaction in the CuO layer.

The solution of the {\sc bcs} equations is predominately of a $d$-wave
character, but because the tetragonal symmetry is broken by the
presence of the chains there is some mixture of $s$-wave order
parameter with no relative phase between the components, although the
relative sign of the order parameter in the planes and chains may be
$\pm 1$. Further, we find that due to a symmetry in the set of coupled
{\sc bcs} equations derived, the sign of the interlayer interaction
affects only the relative sign of the order parameter in the two
layers and not their absolute magnitudes (although some properties
could be affected by this relative sign). We also find that any
interlayer interaction strongly enhances the zero temperature order
parameter, and hence the critical temperature. This is consistent with
the observation that $T_{\rm c}$ is higher in materials with multiple
adjacent CuO$_2$ layers.

Furthermore, we find that the presence of gap nodes in only one of the
layers is enough to produce a low temperature linear behaviour for the
penetration depth, although if the minimum gap in the chains is too
large the $\lambda^{-2}_{xx}(T)$ and $\lambda^{-2}_{yy}(T)$ curves can
cross. Our calculation of the magnetic penetration depth gives a form
that is similar to that measured experimentally perpendicular to the
chains, but not along the chains due to the small value of
$2\mit\Delta_{\rm max}/T_{\rm c}$ in the chains. This leads us to speculate
that there may be intrinsic pairing in the CuO chains of the same
order as in the CuO$_2$ planes since the $2\mit\Delta_{\rm max}/T_{\rm c}$
ratios must both be large (6 to 7) for the low temperature slope of
the $\lambda^{-2}_{ii}(T)$ curves to be approximately equal as is
observed in experiments.\cite{basov,jacobs} This would tend to support
the simple single band orthorhombic model previously proposed by
us.\cite{odonovan3,odonovan2}

Our calculation of the superconducting density of states indicate that
a surface probe may measure very different results depending upon
whether the top layer is CuO chains or CuO$_2$ planes. Depending upon
the interlayer pairing strength the CuO chain layer may have a very
narrow ``$d$-wave'' type gap or a finite ``isotropic $s$-wave'' type
gap.

Our calculation of the $c$-axis Josephson resistance-tunneling
current products, $RJ(T=0)$, for a YBCO-Pb junction for several
choices of $g_{\alpha\beta}$ range from 0.18meV to 0.50meV for the
planes and 2.35meV to 3.06meV for the chains, with possibly a relative
sign between the chain and plane layers due to the relative sign of
the $s$-wave components of the order parameters. The actual $c$-axis
Josephson resistance-tunneling current product for a junction made
with untwinned YBCO would be some weighted average of the plane and
chain results depending upon the relative abundance of chains and
planes in the top layer of the YBCO. For a twinned sample with both
twins equally abundant there would be zero net tunneling current
although there is evidence that for single crystals of YBCO there can
have up to a 5:1 ratio in the relative abundance of the two twin
orientations.\cite{gaulin}

\section*{Acknowledgements}

Research supported in part by the Natural Sciences and Engineering
Research Council of Canada ({\sc nserc}) and by the Canadian Institute
for Advanced Research ({\sc ciar}). We would like to thank W. A.
Atkinson and A. M. Westgate for discussions and insights.


\end{document}